\documentclass[]{aa}
\usepackage[utf8]{inputenc}
\usepackage[T1]{fontenc}
\usepackage{txfonts}
\usepackage{graphicx}
\usepackage{natbib}
\usepackage{hyperref}
\bibpunct{(}{)}{;}{a}{}{,} 

\makeatletter
\renewcommand*\aa@pageof{, page \thepage{} of \pageref*{LastPage}}
\makeatother

\begin{document}

\title{The proper motion of sub-populations in $\omega$~Centauri}
\titlerunning{Motion of sub-populations in $\omega$~Cen}
\author{N.~Sanna\inst{\ref{oaa}}, 
        E.~Pancino\inst{\ref{oaa},\ref{ssdc}},
        A.~Zocchi\inst{\ref{esa}},
        F.~R.~Ferraro\inst{\ref{unibo}},
        P.~B.~Stetson\inst{\ref{dao}} 
        }
\authorrunning{N. Sanna et al.}
\institute{INAF - Osservatorio Astrofisico di Arcetri, Largo E. Fermi 5, 50125 Firenze, Italy\label{oaa}
\and ASI Science Data Center, Via del Politecnico SNC, I-00133 Rome, Italy\label{ssdc}
\and European Space Research and Technology Centre (ESA/ESTEC), Keplerlaan 1, NL-2201 AZ Noordwijk, the Netherlands\label{esa}
\and Dipartimento di Fisica e Astronomia, Università degli Studi di Bologna, via Gobetti 93/2, I-40129 Bologna, Italy\label{unibo}
\and Herzberg Astronomy and Astrophysics, National Research Council, 5071 West Saanich Road, Victoria, British Columbia V9E 2E7, Canada\label{dao}}
   
\date{Received: December 2019}

\abstract{
The galactic globular cluster $\omega$~Centauri is the most massive of its kind, with a complex mix of multiple stellar populations and several kinematic and dynamical peculiarities. Different mean proper motions have been detected among the three main sub-populations, implying that the most metal-rich one is of accreted origin. This particular piece of evidence has been a matter of  debate because the available data have either not been sufficiently precise or limited to a small region of the cluster to ultimately confirm or refute the result. Using astrometry from the second {\em Gaia} data release and recent high-quality, multi-band photometry, we are now in a position to resolve the controversy. We reproduced the original analysis using the {\em Gaia} data and found that the three populations have the same mean proper motion. Thus, there is no need to invoke an accreted origin for the most metal-rich sub-population.}


\maketitle{}


\section{Introduction}

Of all globular clusters (hereafter GCs) in the Milky Way, $\omega$~Centauri ($\omega$~Cen, NGC\,5139) is the most massive  \citep[3.24~10$^6M_{\odot}$,][]{zocchi17} and the most complex in terms of its sub-populations. It is known to host from a minimum of three \citep{pancino00,ferraro04} to at least 15 \citep{bellini17} sub-populations. The complexity is observed in colour-magnitude diagrams (hereafter CMDs), where it appears most clearly in  Hubble Space Telescope (HST) photometry \citep{bellini17}, which shows several co-existing main sequences. The cluster is even more complex from the point of view of its chemistry, with a large spread in metallicity \citep{norris96,pancino02}, extreme multiple populations \citep{gratton11,bastianlardo}, including strong enhancements in helium \citep{dupree13}, and s-process elements \citep{dorazi11}.

The complexity of $\omega$~Centauri is reflected in its kinematics, but often with controversial results. In their study of 400 red giants, \citet{norris97} found that the metal-rich population is more centrally concentrated and kinematically cooler than the metal-poor population \citep[see also][]{sollima07,bellini09b}. Moreover, the metal-poor stars show systemic rotation, while the metal-rich stars seem to be non-rotating. These results were confirmed by \citet{vandeven06} and \citet{bellini18}, who used different sets of data to find differences in the radial distribution and rotation of the sub-populations, as well as possible differences in their anisotropy. However, based on their radial velocity investigations, \citet{pancino07} and \citet{vanloon07} did not find any significant difference in the rotation or velocity spreads among the sub-populations. Another  controversial topic that lingers is the possible presence of an intermediate-mass black hole, initially proposed by \citet{noyola08}, but later put into doubt by, for example,  \citet{vandermarel} or \citet{zocchi19}, who studied the effect of several dynamical ingredients on reproducing the available data.

\begin{figure}[t]
    \centering
    \vspace{-0.3cm}
    \resizebox{\hsize}{!}{\includegraphics[clip]{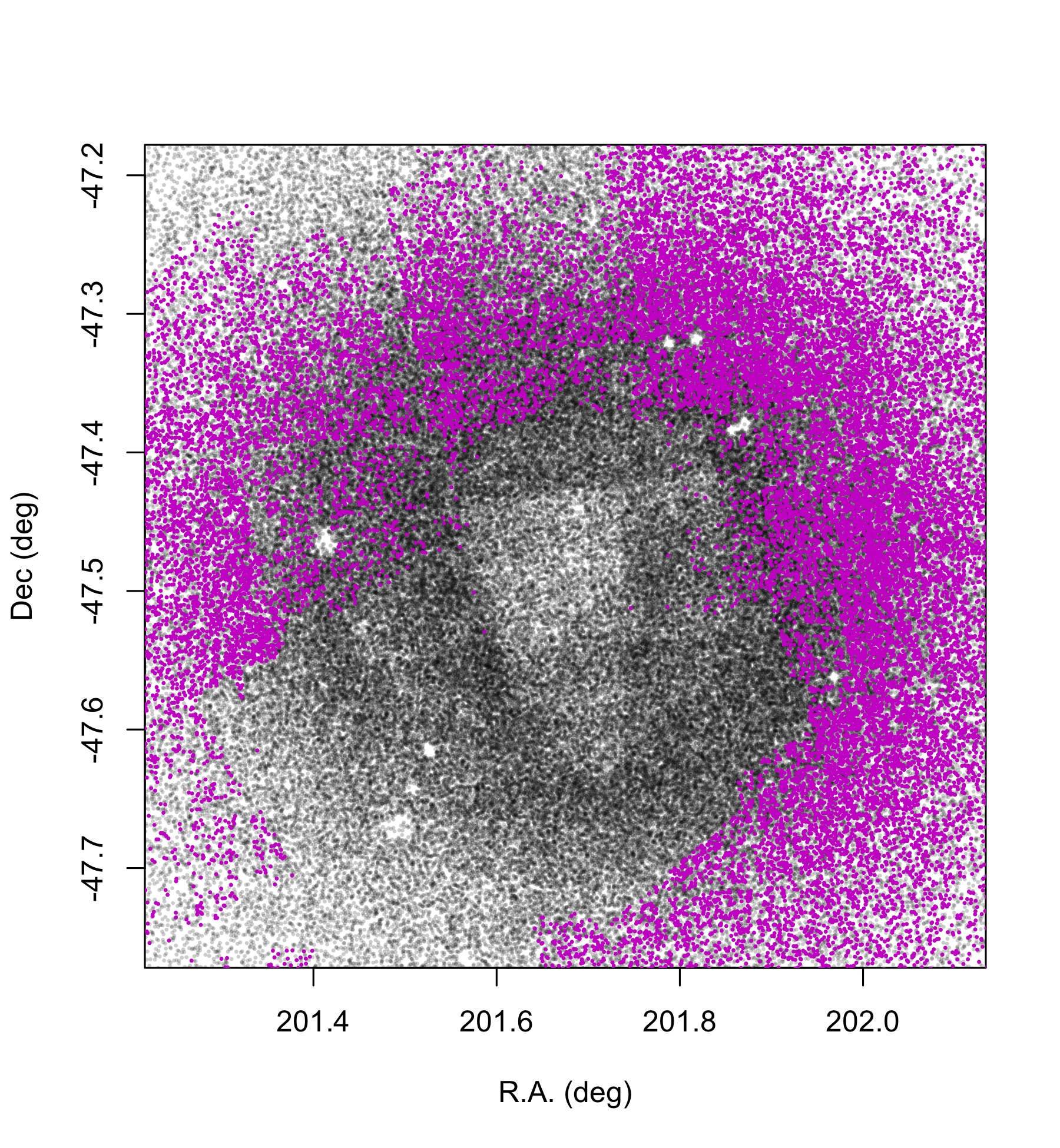}}
    \caption{Map of the central regions of $\omega$~Centauri in the {\em Gaia} DR2 catalogue. The holes caused by bright stars and the irregularly shaped incomplenetess pattern in the central parts can be appreciated. The members by \citet{gaiaGC} are marked as purple dots: their very strict selection leaves a void in the central parts.}
    \label{fig:map}
\end{figure}

The recent second {\em Gaia} data release \citep[hereafter DR2,][]{gaiadr2,gaia16} provided data that could in principle settle some of the open issues, but unfortunately the central regions of $\omega$~Centauri are incomplete, as illustrated in Figure~\ref{fig:map}. The incompleteness is caused by a combination of crowding effects, along with incomplete  {\em Gaia} coverage due to the scanning law, and quality filtering of the catalogue prior to the release. The extremely strict membership selection performed by \citet{gaiaGC} exacerbates the incompleteness, producing a large void in the central parts of the GC \citep[Figure~\ref{fig:map}, see also Figure A.6 by][]{gaiaGC}. The quality is expected to improve significantly with future {\em Gaia} releases \citep{pancino17}. It is not surprising, therefore, that no detailed analysis of the internal kinematics and dynamics of $\omega$~Cen, based on {\em Gaia} DR2 data, has yet appeared: so far, only studies on the systemic properties \citep{bianchini18,sollima19} and on the tidal tails and stream \citep{myeong18,ibata19} have been published. \citet{baumgardt19} derived a velocity dispersion profile for $\omega$~Cen with {\em Gaia} DR2 data, but no study of the kinematic differences among sub-populations in $\omega$~Cen has been published so far. 

There is, however, one dispute about a particularly controversial piece of evidence that can be settled with the present {\em Gaia} DR2 astrometry, even considering the limitations in the case of the central $\omega$~Cen regions. Combining the photometry by \citet{pancino00} with the proper motions by \citet{vanleeuwen00}, \citet{ferraro02} investigated the proper motions of three RGB sub-populations, which they labelled  metal-poor (RGB-MP), metal-intermediate (RGB-MInt), and metal-rich or anomalous (RGB-a or), concluding that the metal-rich sub-population should have an independent origin because its mean proper motion is not consistent with the bulk of the RGB stars. In other words, the RGB-a population would be an accreted system, not yet fully mixed, that was not originally part of the $\omega$~Cen main body; it may be, perhaps, a small GC of the original parent galaxy.

This result has been debated since. \citet{platais03} suggested that it was an artefact caused by instrumental effects because the telescope used to obtain the proper motions by \citet{vanleeuwen00} was moved from South Africa to Australia, so the optics and detectors were not the same in different epochs. \citet{hughes04} showed that a small colour term of 1~mas\,yr$^{-1}$ in the proper motions was properly corrected by \citet{vanleeuwen00} and it was in the opposite sense with respect to the RGB-a motion, thus the catalogue was reliable and the result solid. A deeper discussion of the problem can also be found in \citet{bellini09}, who presented a proper motion investigation using ground-based photometry and astrometry and found no proper motion difference among sub-populations. The authors suggested that an internal stellar proper motions investigation was required, but at that time a catalogue of sufficient quality was not available. More recently, \citet{bellini18} and \citet{libralato18} investigated the proper motions of small external regions of the main sequence (MS) sub-populations of the cluster using HST data and found that all the sub-populations -- which do not correspond exactly to those defined by \citet{ferraro02} -- share the same median proper motions within the uncertainties. Any  difference they found was very small, much smaller than the $\simeq$0.8~mas\,yr$^{-1}$ found by \citet{ferraro02}. However, the HST astrometry was confined to small-area pointings in $\omega$~Cen.

\begin{figure}[t]
    \centering
    \includegraphics[width=1.02\columnwidth,clip]{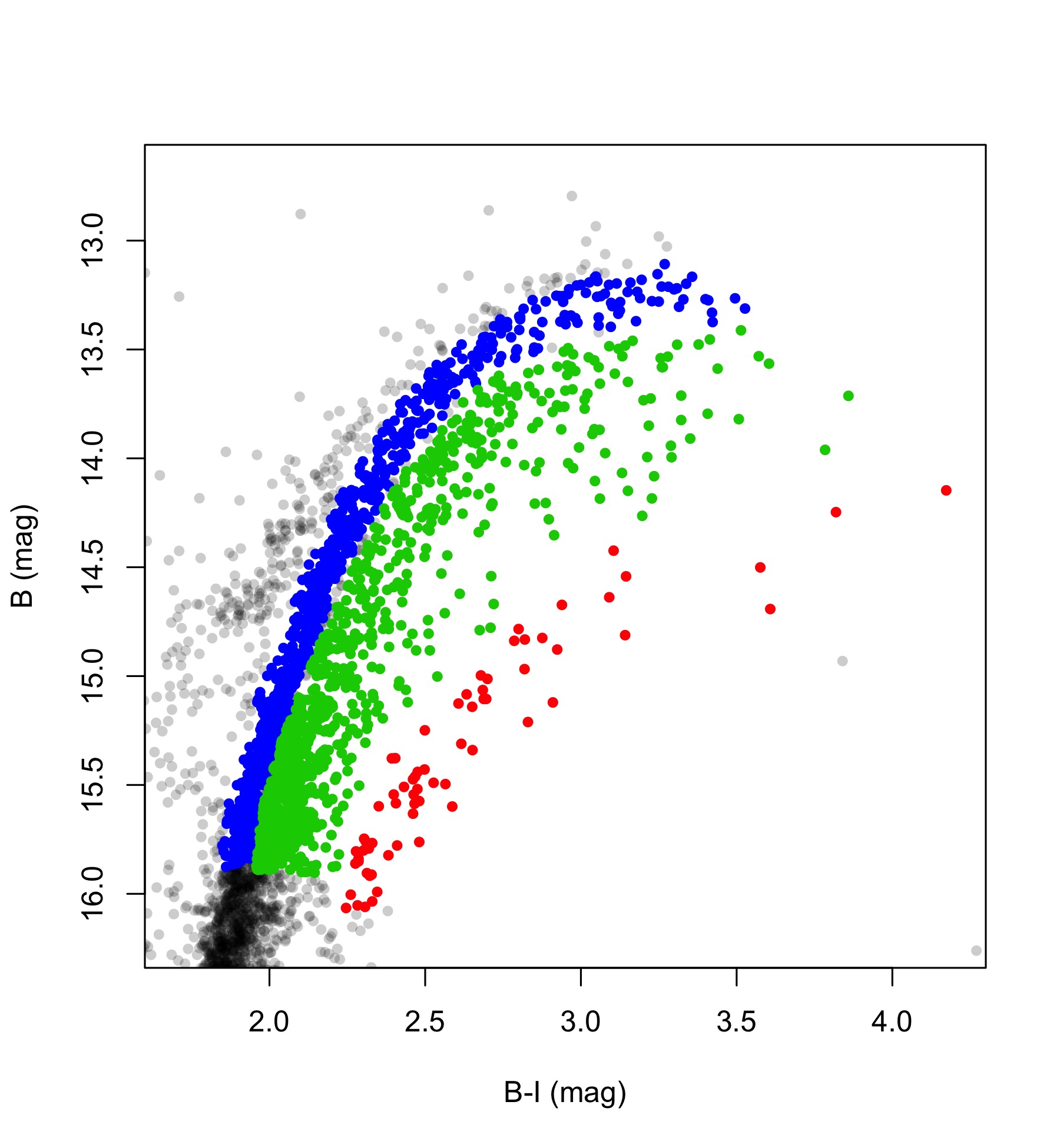}
    \vspace{-0.3cm}
    \caption{Selection of the three sub-populations following criteria as similar as possible to those by \citet{ferraro02}. The grey dots in the background represent the selected members of $\omega$~Centauri; blue dots represent the RGB-MP sample, following the nomenclature by \citet{pancino00} and \citet{ferraro02}; green dots: the RGB-Mint sample; and red dots: the RGB-a sample.}
    \label{fig:cmd}
\end{figure}

Here, we profit from the updated multi-band photometry that was recently published by \citet{stetson19} and the exquisite proper motions available in {\em Gaia} DR2 to revisit the proper motion investigation of the three RGB sub-populations in $\omega$~Centauri. In particular, the {\em Gaia} DR2 astrometry has a much higher quality compared to any previous ground-based catalogue and covers the entire extent of the cluster, unlike previous HST astrometry, and therefore it is the only available astrometric catalogue that has the potential to finally settle this open controversy.


\begin{figure*}[t]
    \centering
    \includegraphics[width=0.85\textwidth,clip]{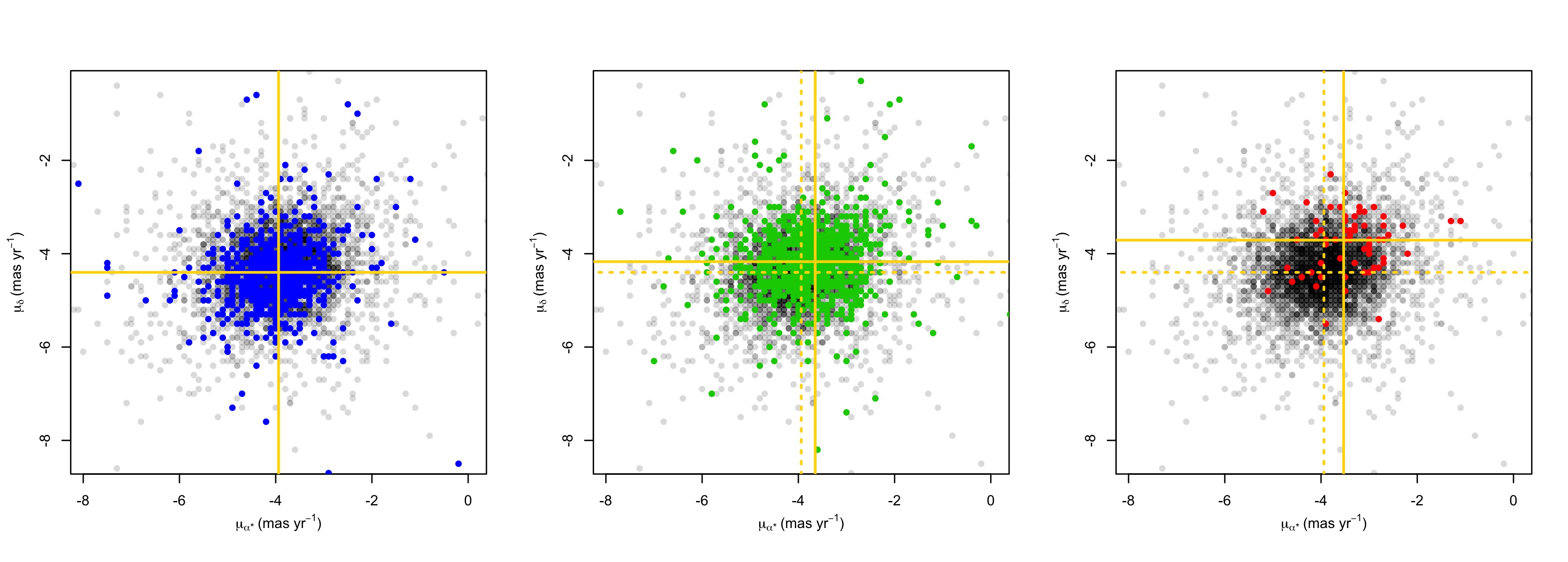}
    \caption{Proper motion vector diagram of the selected members of $\omega$~Centauri (grey dots in all panels) using the astrometry by \citet{vanleeuwen00}. Each panel represents one RGB sub-population with the same colours as in Figure~\ref{fig:cmd}. The solid yellow lines are the mean motion of each sub-population according to \citet{ferraro02}, while the RGB-MP mean motion is reported as a dotted yellow line in the middle and right panels. The RGB-Mint and RGB-a populations are clearly offset from the the RGB-MP one, as was found by \citet{ferraro02}.}
    \label{fig:fvl}
\end{figure*}
\begin{figure*}[t]
    \centering
    \includegraphics[width=0.85\textwidth,clip]{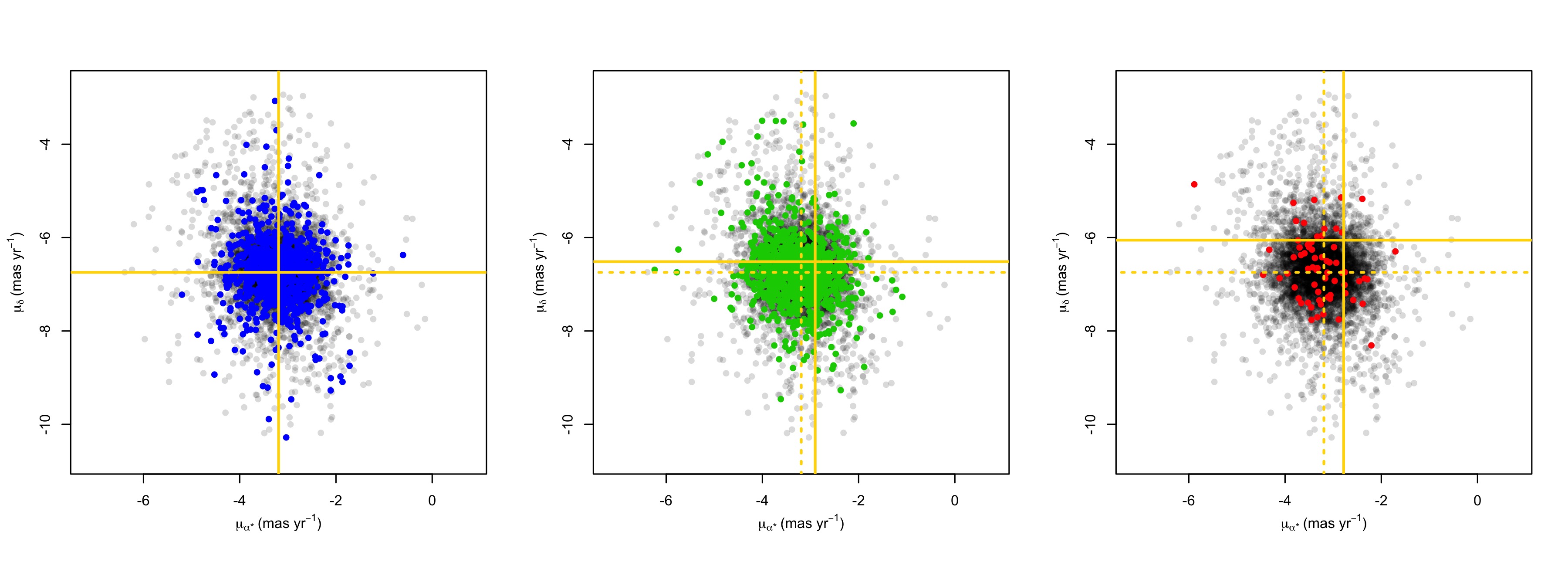}
    \caption{Same as Figure~\ref{fig:fvl}, but using the {\em Gaia} DR2 motions from \citet{gaiaGC}. The yellow lines now correspond to the centroid of the motion by \citet{gaiaGC}, but the offsets of the RGB-Mint and RGB-a from the RGB-MP are the same as in Figure~\ref{fig:fvl}. It is evident that the three populations have compatible mean proper motions.}
    \label{fig:gaia}
\end{figure*}

\section{Data analysis and results}
\label{sec:data}

We based our analysis on the Johnson-Cousins $UBVRI$ photometry by \citet{stetson19} and on the {\em Gaia} DR2 proper motions\footnote{The two catalogues were cross-matched with in-house software by P.~B.~Stetson.}. We also cross-matched the combined Stetson-{\em Gaia} catalogue with the original \citet{vanleeuwen00} astrometry that was used by \citet{ferraro02}\footnote{The cross-match was performed using the CataXcorr package, developed by P.~Montegriffo at the INAF-Osservatorio di Bologna.}. We examined some of the quality flags provided by {\em Gaia} DR2 for this sample, limiting our analysis to the red giant stars with $G$<17~mag. We verified that the behaviour of the astrometric excess noise, goodness of fit, number of good observations used, and RUWE of the selected stars did not deviate from the typical collective behaviour of the sample; thus, we did not apply any specific selection. The initial catalogue of stars in common among the three sources contained 8613 stars. 

\begin{table}
\caption{Comparison of (weighted) mean proper motions for the sub-population samples, among different literature sources, where F02 stands for \citet{ferraro02}, V00 for \citet{vanleeuwen00}, and G18 for \citet{gaiaGC}.}
\label{tab:pm} 
\centering                         
\begin{tabular}{lrrl}        
\hline\hline                
Population & $\mu_{\alpha^*}$ & $\mu_{\delta}$ & Reference \\  
           & (mas\,yr$^{-1}$) & (mas\,yr$^{-1}$) & \\ 
\hline                       
RGB-MP     &  --4.00$\pm$0.02 & --4.44$\pm$0.02  & Here, using V00 \\
RGB-Mint   &  --3.76$\pm$0.03 & --4.20$\pm$0.02 &  \\
RGB-a      &  --3.44$\pm$0.08 & --3.78$\pm$0.08 &  \\
\hline                                   
RGB-MP     &  --3.94 & --4.40 & F02, using V00 \\
RGB-Mint   &  --3.65 & --4.27 &  \\
RGB-a      &  --3.53 & --3.71 &  \\
\hline                                   
RGB-MP     &  --3.21$\pm$0.02 & --6.73$\pm$0.02 & Here, using G18 \\
RGB-Mint   &  --3.27$\pm$0.02 & --6.74$\pm$0.02 &  \\
RGB-a      &  --3.28$\pm$0.07 & --6.64$\pm$0.07 &  \\
\hline                                   
\end{tabular}
\end{table}

We could not use the membership selection by \citet{gaiaGC} because it is quite restrictive and it does not contain enough stars in the RGB-a population (see also Figure~\ref{fig:map}). Therefore, we performed a less restrictive membership selection, with the goal of cleaning the catalogue from obvious non-members, by retaining all stars within a 5D ellipsoid defined as follows:

\begin{displaymath}
\frac{(\alpha-\alpha_0)^2}{r^2} + \frac{(\delta-\delta_0)^2}{r^2} + \frac{(\mu_{\rm{\alpha^*}}-\bar{\mu}_{\rm{\alpha^*}})^2}{(3\sigma_{\mu_{\rm{\alpha^*}}})^2} + 
\frac{(\mu_{\rm{\delta}}-\bar{\mu}_{\rm{\delta}})^2}{(3\sigma_{\mu_{\rm{\delta}}})^2} + \frac{(\varpi-\bar{\varpi})^2}{(5\sigma_{\varpi})^2} < 1,
\end{displaymath}
where $r$=30$^{\prime}$ (the maximum extension allowed for a uniform coverage in the ground-based photometry); ($\alpha_0$,$\delta_0$) are the central coordinates of $\omega$~Centauri by \citet{stetson19}; ($\bar{\mu}_{\rm{\alpha^*}}$,$\bar{\mu}_{\rm{\delta}}$,$\bar{\varpi}$) are the systemic motion and parallax measurements by \citet{gaiaGC}\textbf{ \footnote{As customary,  $\mu_{\rm{\alpha^*}}=\mu_{\alpha}~\cos{\delta}$ \citep[see, e.g.][]{lindegren16}.}}; ($\sigma_{\mu_{\rm{\alpha^*}}}$,$\sigma_{\mu_{\rm{\delta}}}$,$\sigma_{\varpi}$) represent the median absolute deviation (MAD) of the proper motion and parallax distributions of members, refined after a few iterations, and set to (1.09,1.29)~mas\,yr$^{-1}$ and 0.69~mas, respectively. Following the selection, the sample was made up of 5113 stars. The typical (median) uncertainties on the individual stars are in the range of 0.08--0.12~mas~yr$^{-1}$ in the case of $\mu_{\rm{\alpha^*}}$ and 0.12--0.17~mas~yr$^{-1}$ in the case of $\mu_{\rm{\delta}}$. We note that the errors are larger for the bright stars (G$\lesssim$13~mag) than for the faint ones, owing to the fact that {\em Gaia} is not designed to target bright stars.

\begin{figure*}[t]
    \centering
    \includegraphics[width=\columnwidth]{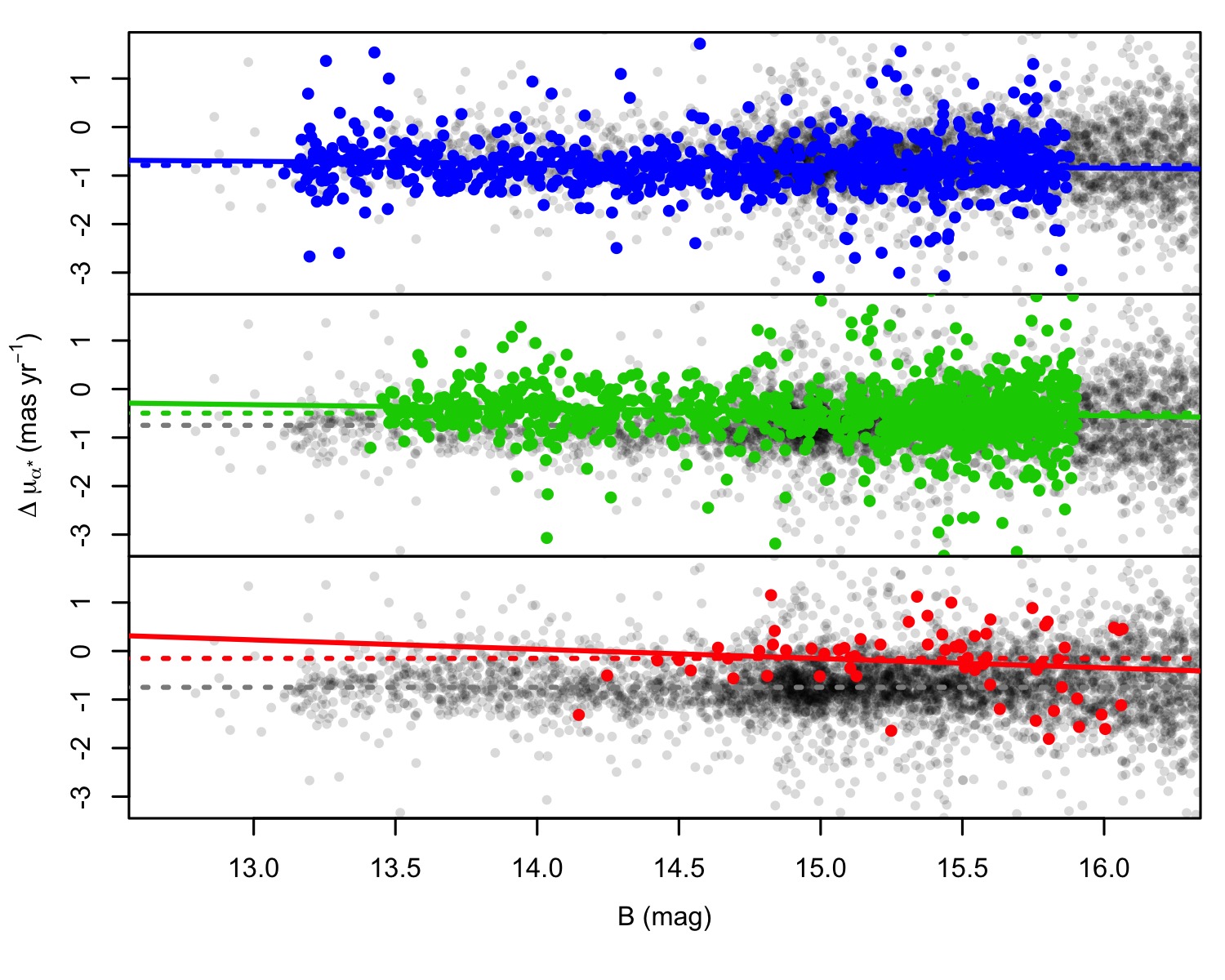}
    \includegraphics[width=\columnwidth]{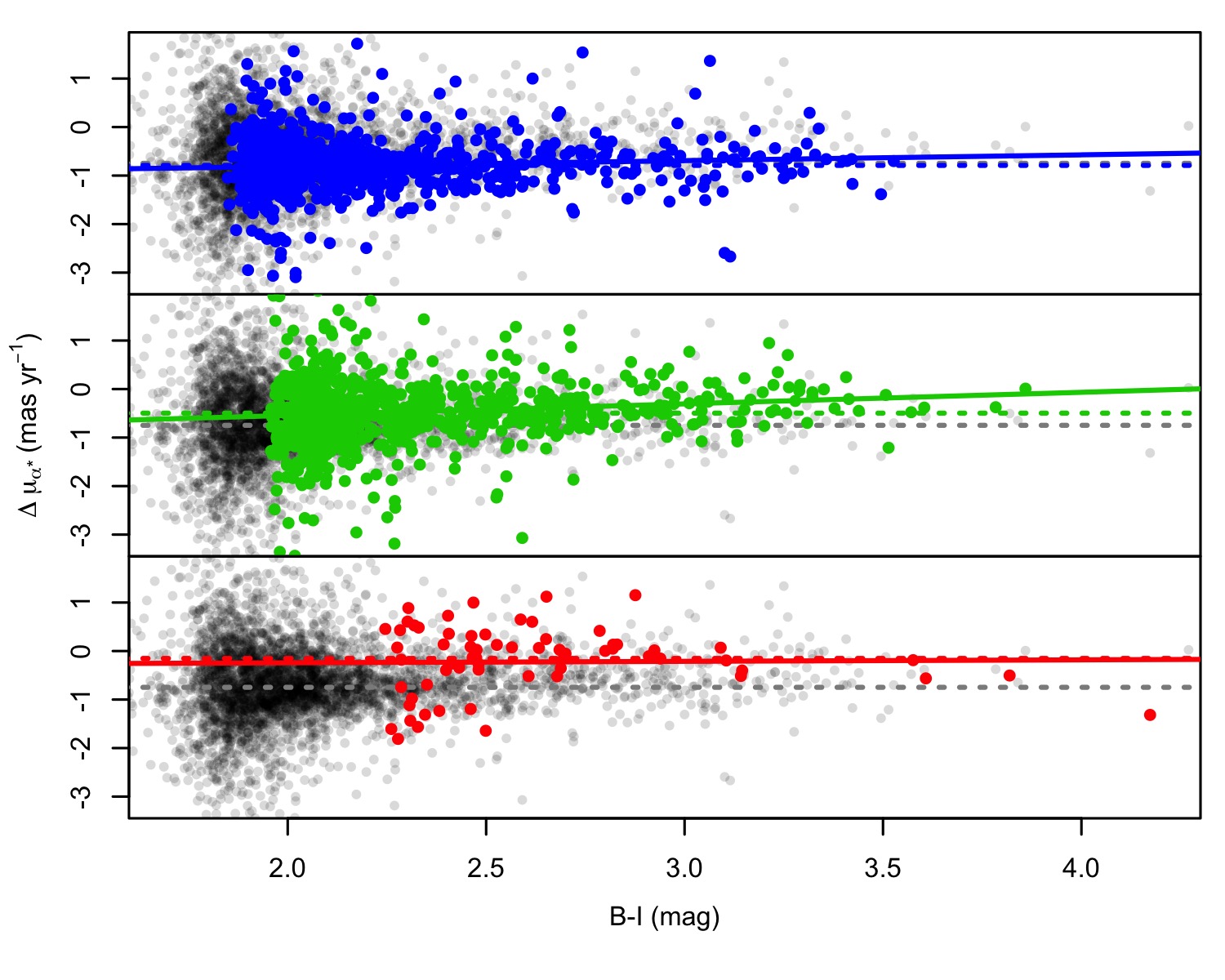}
    \includegraphics[width=\columnwidth]{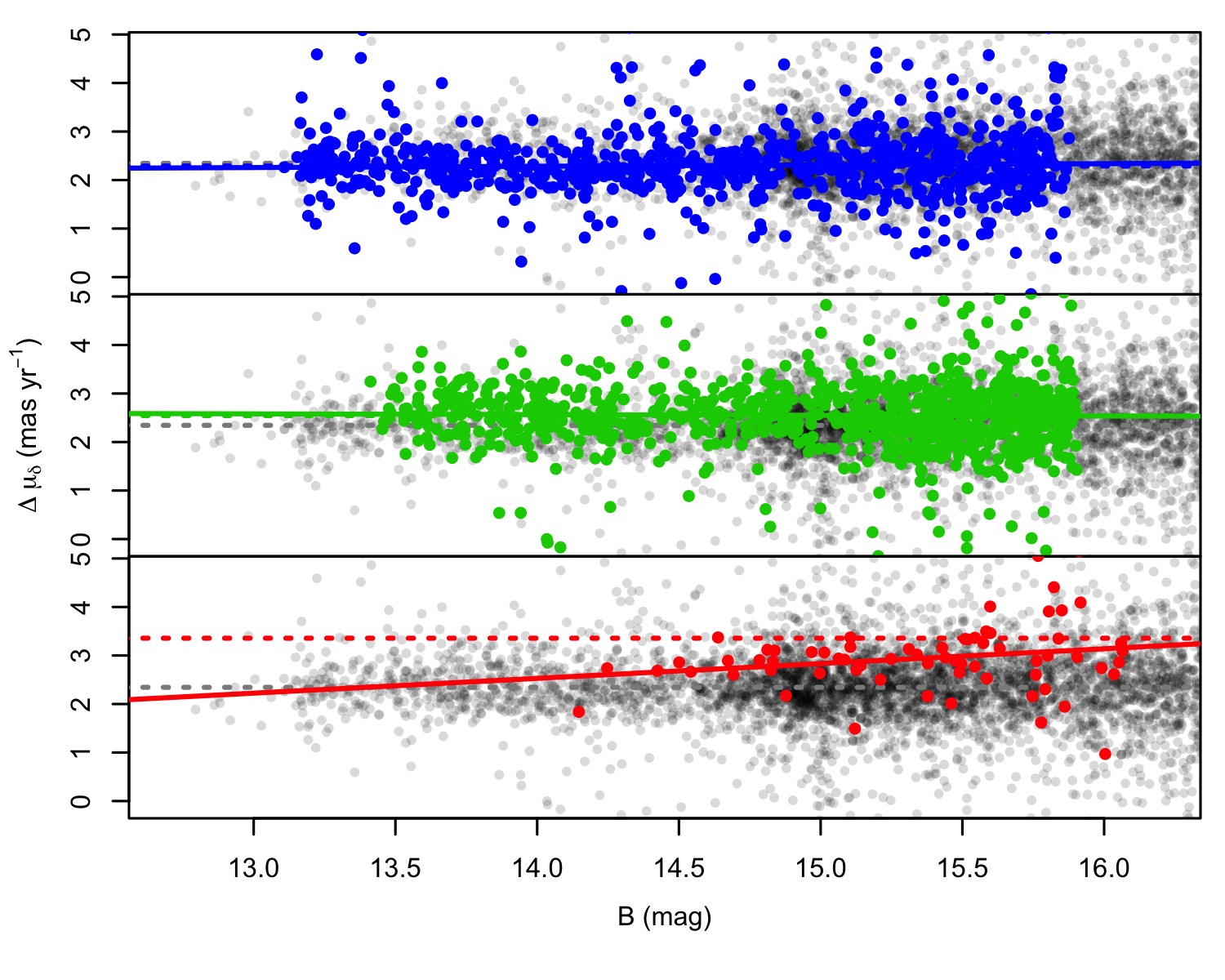}
    \includegraphics[width=\columnwidth]{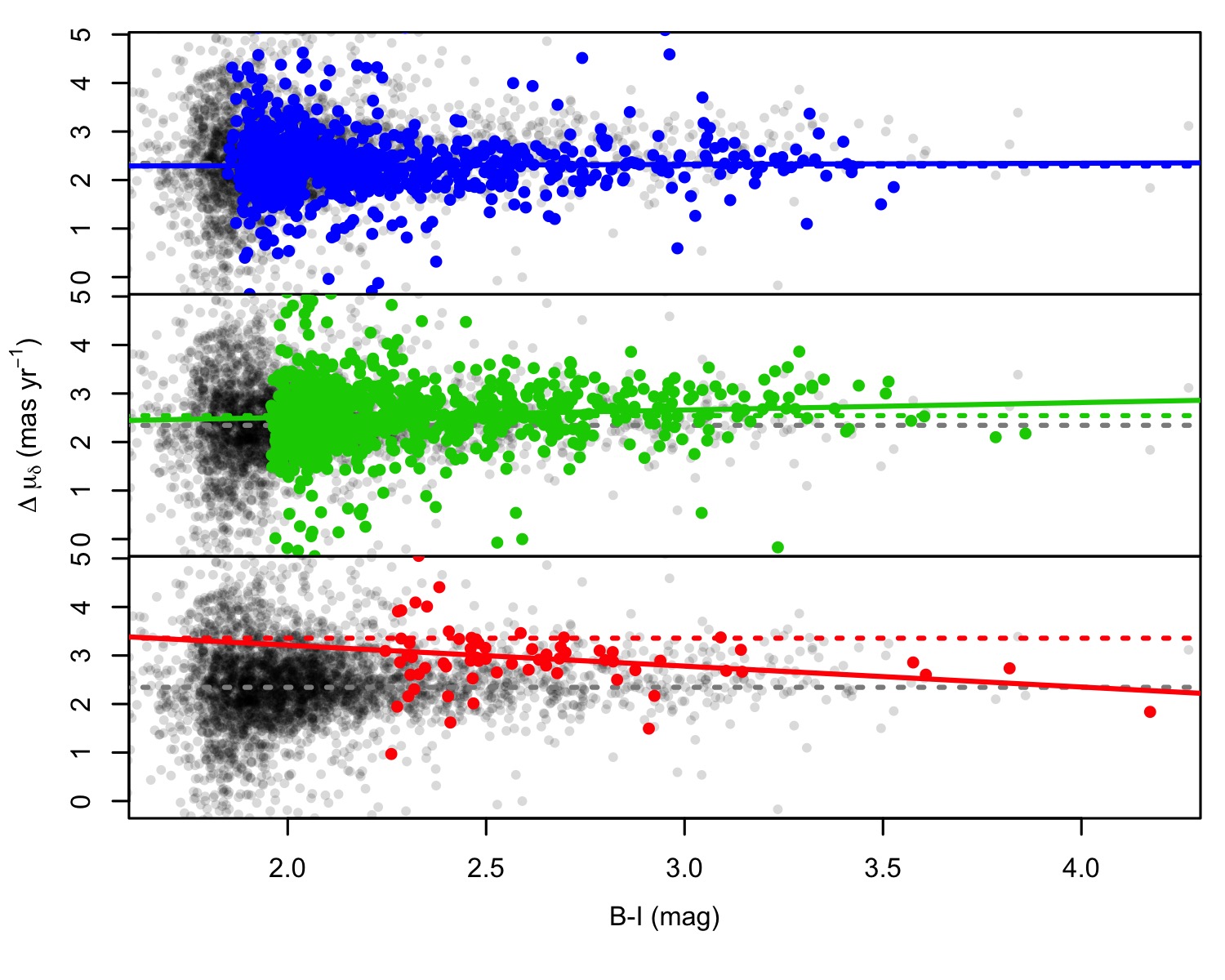}
    \caption{Difference of the proper motion in RA (top panels) and Dec (bottom panels) by \citet{vanleeuwen00} and \citet{gaiaGC} as a function of magnitude (left panels) and of colour (right panels). Stars are coloured as in the previous figures. The average difference between the two catalogues are plotted as grey dotted lines, the ones for each sub-population are plotted as dotted lines of the respective colours, and the linear fits for each sub-population are shown as solid lines of the respective colours. None of the slopes are statistically significant, even the few that appear large are driven by single data points and a small sample size (see Section~\ref{sec:data} and Table~\ref{tab:fit}).}
    \label{fig:comp}
\end{figure*}

We manually selected  the three RGB sub-population samples following the criteria by \citet{ferraro02} as closely
as possible, as shown in Figure~\ref{fig:cmd}. In particular, we used the same limiting magnitudes for the selection of the populations labelled by \citet{pancino00} and \citet{ferraro02} as RGB-MP and RGB-MInt, while for the so-called RGB-a population we selected slightly fainter stars (B$\lesssim$16.1~mag) thanks to the clearer separation from the bulk of the RGB population. As a first sanity check, adopting our selections, we reproduced the original result by \citet{ferraro02} using the \citet{vanleeuwen00} astrometry to make sure that our selection of the three sub-populations was comparable with theirs. The result is shown in Figure~\ref{fig:fvl}, where we do indeed observe that the populations labelled by \citet{pancino00} and \citet{ferraro02} as RGB-MInt and RGB-a have a different mean proper motion compared to the RGB-MP population. In particular, as shown in Table~\ref{tab:pm}, our results are fully compatible with the offsets found by \citet{ferraro02}. We then repeated the same experiment with the {\em Gaia} DR2 proper motions, as illustrated in Figure~\ref{fig:gaia} and with the results reported in Table~\ref{tab:pm}, but in this case all three populations appear clearly compatible with the same mean proper motion and with the \citet{gaiaGC} systemic value for the cluster. We note here that the systemic motion of $\omega$~Cen derived by \citet{vanleeuwen00} and by \citet{gaiaGC} are quite different from each other, especially as far as $\mu_{\delta}$ is concerned. Both estimates are quite different from the one by \citet{dinescu99} as well.

\begin{table*}
\caption{Linear fits and correlation coefficients for the sub-populations (see Figure~\ref{fig:comp}). For each population, a linear fit of $y$ (the proper motion component) as a function of $x$ (magnitude or colour) is performed. The angular coefficent $m$ and the intercept $q$ of each fit are listed, along with the statistical significance of the fit  $p$. The correlation coefficients by Pearson ($r$), Spearman ($\rho$), and Kendall ($\tau$) are also reported (see text for a discussion).}
\label{tab:fit} 
\centering                         
\begin{tabular}{lllrrrrrr}        
\hline\hline                
Population & $x$ & $y$ & $m$ & $q$ & $p$ & $r$ & $\rho$ & $\tau$ \\ 
\hline                       
RGB-MP   & B   & ${\mu}_{\rm{\alpha^*}}$ & --0.10 $\pm$ 0.54 & --0.046 $\pm$ 0.037 & 0.21 & --0.044 & --0.005 & --0.005 \\
         & B   & ${\mu}_{\rm{\delta}}$   &   1.91 $\pm$ 0.59 &   0.027 $\pm$ 0.040 & 0.51 &   0.023 &   0.021 &   0.013 \\
         & B-I & ${\mu}_{\rm{\alpha^*}}$ & --1.05 $\pm$ 0.19 &   0.119 $\pm$ 0.086 & 0.17 &   0.048 & --0.001 &   0.001 \\
         & B-I & ${\mu}_{\rm{\delta}}$   &   2.25 $\pm$ 0.21 &   0.024 $\pm$ 0.094 & 0.80 &   0.009 & --0.008 & --0.006 \\
\hline                                   
RGB-MInt & B   & ${\mu}_{\rm{\alpha^*}}$ &   0.67 $\pm$ 0.83 & --0.076 $\pm$ 0.055 & 0.17 & --0.044 & --0.101 & --0.070 \\
         & B   & ${\mu}_{\rm{\delta}}$   &   2.77 $\pm$ 0.62 & --0.015 $\pm$ 0.041 & 0.72 & --0.012 & --0.062 & --0.042 \\
         & B-I & ${\mu}_{\rm{\alpha^*}}$ & --1.02 $\pm$ 0.27 &   0.238 $\pm$ 0.117 & 0.04 &   0.066 &   0.184 &   0.124 \\
         & B-I & ${\mu}_{\rm{\delta}}$   &   2.20 $\pm$ 0.20 &   0.153 $\pm$ 0.087 & 0.08 &   0.057 &   0.150 &   0.101 \\
\hline                                   
RGB-a    & B   & ${\mu}_{\rm{\alpha^*}}$ &   2.72 $\pm$ 3.89 & --0.192 $\pm$ 0.253 & 0.45 & --0.093 & --0.048 & --0.023 \\
         & B   & ${\mu}_{\rm{\delta}}$   & --1.74 $\pm$ 2.69 &   0.305 $\pm$ 0.175 & 0.09 &   0.212 &   0.212 &   0.159 \\
         & B-I & ${\mu}_{\rm{\alpha^*}}$ & --0.30 $\pm$ 0.85 &   0.031 $\pm$ 0.321 & 0.92 &   0.012 &   0.029 &   0.006 \\
         & B-I & ${\mu}_{\rm{\delta}}$   &   4.07 $\pm$ 0.58 & --0.431 $\pm$ 0.219 & 0.05 & --0.237 & --0.185 & --0.143 \\
\hline                                   
\end{tabular}
\end{table*}

As a final check, we plot the star-by-star differences of the \citet{vanleeuwen00} and \citet{gaiaGC} proper motions as a function of magnitude and colour (Figure~\ref{fig:comp}). We checked for the presence of significant slopes using four indicators: the p-values of the angular coefficient of linear fits, the Spearman rank coefficient $\rho$, the Pearson correlation coefficient $r$, and the Kendall rank coefficient $\tau$  (Table~\ref{tab:fit}). We found no significant slopes in any of the samples against colour or magnitude, contradicting the finding by \citet{platais03} but confirming the findings by \citet{tesi} and \citet{hughes04} in this respect. Even the apparent slopes visible in some of the panels of Figure~\ref{fig:comp} are not significant and driven by single data points and sample sizes. As can be seen in Table~\ref{tab:fit}, the errors on the angular coefficents $m$ and intercepts $q$ of the fits are very large. Besides the p-value of the linear fit, all three correlation tests have also large p-values (generally well above $\simeq$0.05) and the correlation coefficients $\rho$, $r$, and, $\tau$ are always much closer to zero than to $\pm$1.

In past studies, the absence of a significant residual slope of the proper motion as a function of colour and magnitude was taken as proof that no spurious effect was present in the \citet{vanleeuwen00} catalogue. However, from Figure~\ref{fig:comp}, it is evident that this condition was necessary but not sufficient; even if no slope is present in the data, each sub-population clearly drifts away from the mean motion of the cluster in one catalogue but not in the other. We can also see from Figure~\ref{fig:comp} that the offsets found in Figure~\ref{fig:fvl} and Table~\ref{tab:pm} are entirely compatible with the differences between the proper motion measurements in the two catalogues, suggesting a spurious measurement effect in the \citet{vanleeuwen00} catalogue. 


\section{Conclusions}

Our main result is  that when using {\em Gaia} DR2 data, the three sub-populations have compatible mean motions with each other and with the \citet{gaiaGC} systemic motion of $\omega$~Cen. Previous astrometric catalogues, most notably those by \citet{bellini09} and \citet{bellini18}, have provided the same result. However, \citep{bellini09} could count on astrometric errors of the order of a few mas\,yr$^{-1}$, that is, a few times larger than the putative proper motion offsets among sub-populations, similarly to \citet{vanleeuwen00}, while the HST astrometric catalog by \citet{bellini18} had sub-mas uncertainties, comparable to those in the {\em Gaia} DR2 catalogue but, of course, limited to a very small area. Thus, {\em Gaia} DR2 is the only presently available catalogue with sufficient quality and area coverage to settle the controversy.

The conclusion that can be drawn from the available literature body and the present analysis is that, indeed, as was suggested by \citet{platais03}, the \citet{vanleeuwen00} catalogue contained spurious instrumental effects, although they were not so immediately evident as a simple colour or magnitude trend. Indeed, a colour trend of about 1~mas\,yr$^{-1}$ was found and removed from the \citet{vanleeuwen00} catalogue, as pointed out by \citet{hughes04}, but this was not sufficient to correct the problem. This implies that it is not necessary to assume that the RGB-a population was an external system which was then accreted by the main body of $\omega$~Cen. We expect that the next {\em Gaia} releases will include a treatment of crowding effects \citep{pancino17} and will rely on more high-quality data in the central parts. Combined perhaps with HST astrometry, {\em Gaia} data do have the potential to help us decipher the complex kinematic structure and evolution of $\omega$~Cen and its sub-populations.


\begin{acknowledgements}
We thank the referee for the useful suggestions. NS and EP acknowledge the financial support to this research by INAF, through the Mainstream Grant 1.05.01.86.22 assigned to the project “Chemo-dynamics of globular clusters: the Gaia revolution” (P.I. E. Pancino). AZ acknowledges support through a ESA Research Fellowship. This work has used data from the European Space Agency (ESA) mission \textit{Gaia} (\url{https://www.cosmos.esa.int/gaia}), processed by the \textit{Gaia} Data Processing and Analysis  Consortium (DPAC, \url{https://www.cosmos.esa.int/web/gaia/dpac/consortium}). Funding for the DPAC  has  been  provided  by  national  institutions, in particular the institutions participating in the \textit{Gaia} Multilateral Agreement. All figures and the statistical analysis were done with the R programming language \citep{R,data.table} and Rstudio (https://www.rstudio.com/).
\end{acknowledgements}


\bibliographystyle{aa} 
\bibliography{omega} 

\end{document}